# Roebel cables from *RE*BCO coated conductors: a one-century-old concept for the superconductivity of the future

Wilfried Goldacker[1], Francesco Grilli[1], Enric Pardo[2], Anna Kario[1], Sonja I. Schlachter[1], Michal Vojenciak[2]

[1] Karlsruhe Institute of Technology, Institute for Technical Physics, Hermann-von-Helmholtz-Platz 1, 76344 Eggenstein-Leopoldshafen, Germany

[2] Slovak Academy of Sciences, Institute of Electrical Engineering, Dubravska Cesta 9, 84104 Bratislava, Slovakia

Author to whom correspondence should be addressed: francesco.grilli@kit.edu

**Abstract**

Energy applications employing high-temperature superconductors (HTS), such as motors/generators, transformers, transmission lines and fault current limiters, are usually operated in the alternate current (AC) regime. In order to be efficient, the HTS devices need to have a sufficiently low value of AC loss, in addition to the necessary current-carrying capacity. Most applications are operated with currents beyond the current capacity of single conductors and consequently require cabled conductor solutions with much higher current carrying capacity, from a few kA to up to 20-30 kA for large hydro-generators.

A century ago, in 1914, Ludwig Roebel invented a low-loss cable design for copper cables, which was successively named after him. The main idea behind Roebel cables is to separate the current in different strands and to provide a full transposition of the strands along the cable direction. Nowadays, these cables are commonly used in the stator of large generators. Based on the same design concept of their conventional material counterparts, HTS Roebel cables from *RE*BCO coated conductors were first manufactured at the Karlsruhe Institute of Technology (KIT) and have been successively developed in a number of varieties that provide all the required technical features such as fully transposed strands, high transport currents and low AC losses, yet retaining enough flexibility for a specific cable design. In the past few years a large number of scientific papers have been published on the concept, manufacturing and characterization of such cables. Times are therefore mature for a review of those results. The goal is to provide an overview and a succinct and easy-to-consult guide for users, developers, and manufacturers of this kind of HTS cables.

## 1. Introduction

High-temperature superconductors (HTS) in the form of *RE*BCO (*RE*Ba$_2$Cu$_3$O$_x$ with *RE*=Rare Earth elements) coated conductor (CC) tapes – the second generation of HTS materials – show great potential for use in many applications such as power transmission cables, motors, generators, fault current limiters, transformers and magnets [1]K. Today the superconducting *RE*BCO material is available as commercial product in long lengths with high current-carrying capacity from several providers [2–5]. Implementation of superconductors is particularly attractive for large-scale applications (such as those mentioned above) in virtue of the significantly reduced size and weight (typically a factor 2-3) and increased energy efficiency. The requested current capacity of many devices, however, is significantly beyond that of a single CC tape and demands for assembled high-current cables. In magnets, for example, a higher drive current allows reduction of the winding number for a given field resulting in a lower inductance, which is mandatory for many applications. Most of the devices are operated in alternate current (AC) regime and keeping the occurring AC losses under an acceptable threshold is a quite general issue and a key point for the success of a superconducting solution.

The technological request for low-loss high-current conductors rose up already in very early times when conventional electrical machinery was being scaled to bigger size. At the beginning of the $20^{th}$ century the size of electrical power generators, for example, was limited by the occurring AC losses. The invention of a low-loss assembled copper cable design with insulated strands and full transposition by Ludwig Roebel at the BBC company in Mannheim (Germany) (German patent 1914, see Figure 1), was the breakthrough for AC loss reduction. It paved the way to a new power class of generators, and has established this new design of a high-current low-loss conductor as a standard until present [6]. Roebel was the first person to understand and identify the need of segmenting the conductor into strands (insulated from each other) and of transposing the strands along the cable direction to reduce induced eddy currents and current loops. Nowadays Roebel bars are the standard conductors in the stator windings of large conventional generators and motors [7,8]

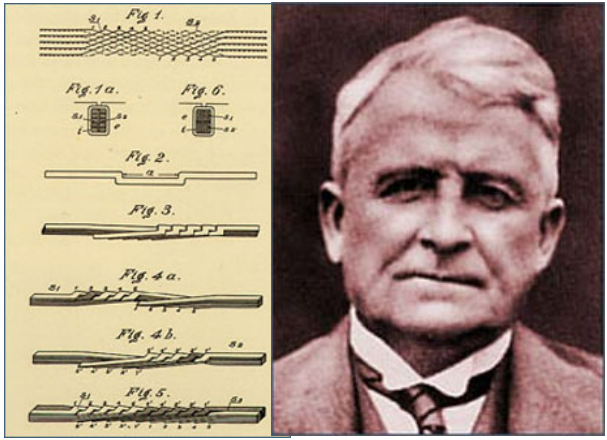

Figure 1. The patent for a low loss cable for generators (1914, BBC Company Mannheim, Germany) and a picture of Ludwig Roebel (https://www.mannheim.de/wirtschaft-entwickeln/roebelstab-ludwig-roebel-1878-1934).

In the field of applied superconductivity, AC currents or ramped fields lead quite soon to the same request to reduce the AC losses by means of a suitable conductor structure with the same features mentioned above – a strand structure and transposition. In the Large Coil Task (LCT) [9], investigated in the years 1980-85, six large superconducting fusion magnets were designed, built and tested for the prospected use in a torus arrangement of a Tokamak fusion reactor. In one unit, the EURATOM magnet, the first superconducting Roebel cable was made from transposed NbTi strands [9], see Figure 2. The ductility of NbTi allowed perfect shaping of the Roebel structure with a quite sharp step-over-bending edge and a quite short transposition length (280 mm). In contrast to the traditional Roebel geometry with narrow but high conductor stacks, this cable was small in height and broad, resembling a Rutherford cable. Cables of this type were used in the dipole magnets of the accelerator facilities at CERN [10]. Due to the planar tape geometry of HTS coated conductors, the required tight bends for such cable designs are impossible in the in-plane direction and new solutions were required.

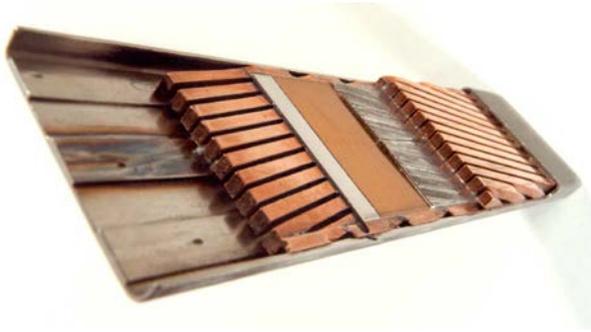

**Figure 2. The NbTi Roebel cable of the EURATOM toroidal field magnet in the Large Coil Task project** [9].

## 2. Cable design and preparation

In this section first the evolution of HTS Roebel cables is summarized; then details about cable design and production methods are discussed; finally several possible modifications of the basic Roebel cable design and alternative concepts for high-current cables are examined.

### 2.1. The evolution of HTS Roebel cables

The first HTS Roebel cable was demonstrated by the Siemens Corporate Technology group using BSCCO(2223) tapes as strands [11]. This cable consisted of 13 strands and showed a self-field reduced DC transport current of about 400 A at 77 K. The quite long transposition length of typically 3 m in such cables was caused by the limited in-plane bending ability of the material. The cable achieved the expected reduced AC losses in comparison with stacked tapes and served as a motivation for the application in HTS transformers [12]. The idea of a Roebel structure with *RE*BCO coated conductor was proposed by Martin Wilson in 1997 [13] at the CEC/ICMC conference, but a way to manage the bending, the strand shape forming and the assembling issue was still unknown. The solution was found a few years later thanks to the progress in homogeneity and robustness of CCs and to their commercial availability. The meander-like shaping of the CC as shown in the CAD model of Figure 3 and presented by Wilfried Goldacker in 2005 [14] was first realized by means of precision punching technique at Forschungszentrum Karlsruhe (now Karlsruhe Institute of Technology, KIT) and published in 2006 [15]. The first cable was made from 16 punched strands of DyBCO coated conductors (manufactured by the THEVA company) and carried a self-field transport current of 500 A at 77 K, about half of the sum of the critical currents of the individual tapes. Unfortunately the cable had no copper stabilization and burned through due to the lack of quench protection. The second presented Roebel cable from 12 mm-wide Cu-stabilized SuperPower MOCVD coated conductor tape already achieved a DC transport current of 1.02 kA at 77 K in self-field. The 0.36 m long sample (two transposition lengths) already showed the reliability of the strand punching process and excellent homogeneity, thanks to the use of coated conductor from SuperPower [16]. Small resistive contributions in the voltage-current (*V-I)* characteristics above $I_c$/2 indicated current redistributions between the strands along the cable. The decrease of the measured transport current by 60% compared to the sum of the critical currents of the individual strands was ascribed to the generated self-field, whose pattern was calculated by means of a numerical method based on a Biot-Savart-approach.

A second group at Industrial Research Limited (IRL) adopted the topic of coated conductor Roebel cables as main subject of investigation, immediately focusing on the industrial production of such cables with machines specifically designed to produce long lengths in an automated fabrication process [17].

Roebel cables from CCs offer a large variety of choices to increase the current, to optimize the cable design and to increase the thermal and mechanical stability by modifying the structure of the composing strands. The potential for further increased transport currents was shown by cabling 3-fold stacks of strands into a 45-strand cable of 1.1 m length and 18.8 cm transposition length carrying 2.6 kA (77 K, self-field) [18]. Increasing the transposition length to be able to add more strands is another possibility to increase the current performance in a simple way.

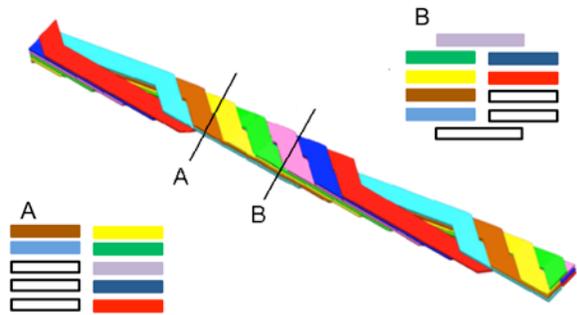

**Figure 3. Schematic illustration of a Roebel bar made from coated conductor tapes. Two transversal cross-sections at different positions are also shown.**

### 2.2. Design parameters

The Roebel cable is made from meander shaped CC tapes. For the Roebel cable design we follow the nomenclature introduced by IRL [19]. The fundamental geometrical parameters are the original tape width $W_T$, the transposition length $L_{TRANS}$ (which is a full period of the meander), the strand width $W_R$, the strand-edge clearance $W_C$ (clearance in the cable centre), the crossover width $W_X$, the inter-strand gap $L_{ISG}$, the crossover angle $\phi$, the cut-off fillet radius $R$, and the inner radius $R_i$. The parameters are shown in the schematic sketch of Figure 4.

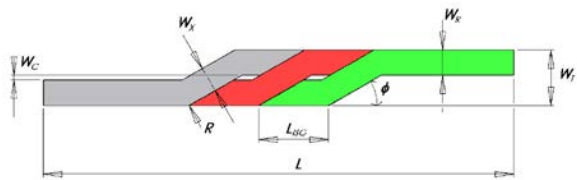

**Figure 4. Nomenclature to specify the geometry of a Roebel cable following reference** [19]. **The inner radius $R_i$ opposite of $R$ is not indicated in this figure, but it is important to manage the stresses at this point, see** [20] **for details.**

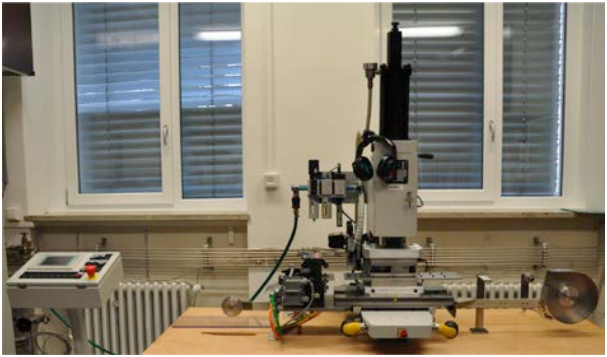

**Figure 5.** Reel-to-reel computer controlled pneumatic punching machine utilized at KIT for producing Roebel strands. The mechanical punching tools can be changed for obtaining cable widths between 4 and 12 mm. At present the transposition length for 4 mm-wide tape is 115.7 mm, whereas for 12 mm-wide tape it is 126 mm.

The transposition length is a suitable parameter to determine the number of applicable strands and needs for a specific application. The inter-strand gap is strongly correlated to the assembling procedure, being narrower for hand-made cables (KIT) and wider in the case of fully automatic process (IRL). Numerical calculations on the geometry with the best mechanical performance and current capacity [21] confirmed the set of parameters usually used at KIT: an inner radius $R_i$ = 2 mm (the stress hot spot in the tape), a strand-edge clearance of 1-2 mm for $W_T$ = 10-12 mm and no clearance for $W_T$ =4 mm (which needs a sufficient inter-strand gap) and a crossover angle of 30 degrees. According to the calculations presented in [21], a sharp outer edge of the crossover section and an increase of the inner radius leads to a reduction of the von Mises stresses under tensile loads. In the Roebel cables from IRL some parameters were chosen differently, in order to be better compatible with the assembling machinery, namely: a larger clearance width $W_c$ between the two sides of the cable and a smaller number of strands per cable length unit [17]. While producing one single strand from a CC tape, the crossover width $W_X$ and the angle can be adjusted for best current performance. If several strands are cut in parallel from a wider CC material (e.g. 40 mm), $W_X$ is predetermined by geometrical reasons and the cross over section becomes the current-limiting part [17].

### 2.3. Production methods

Several methods were investigated for the meander-like shaping process of the CC tapes. Laser cutting with conventional equipment is a very flexible technique, but failed due to heating and melting defects. The newer picoseconds-infrared laser technology avoids producing defects, but it not very economically attractive in terms of production speed. Mechanical punching was identified both by KIT and IRL as the best method and can achieve a precision of <50 micron for the strand width. Figure 5 shows the reel-to-reel punching tool in use at KIT, which allows using the full tape width and varying the transposition length in a wide range. Advantages of this technique are the possibility of adjusting the tool to the tape conditions and its high production speed, typically 50 meters of tape per hour. The loss in current carrying capability from punching can be limited to <2-3% for homogenous CC tapes [16]. Figure 6 shows 10 punched strands before and after they are assembled into cable.

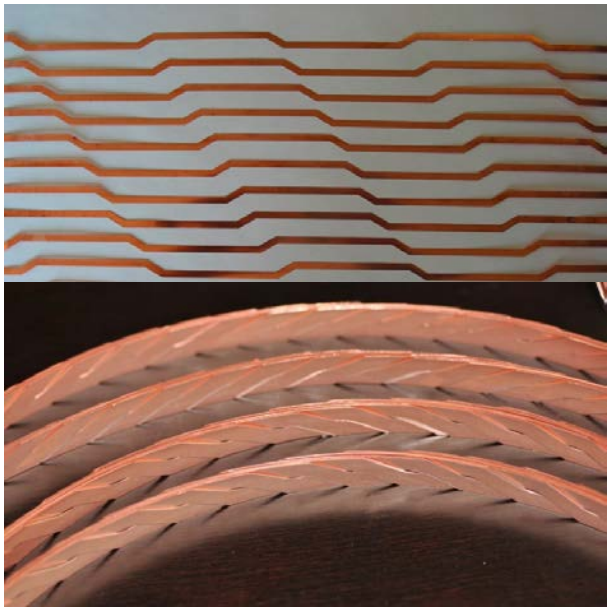

**Figure 6. Ten strands punched from 12 mm-wide coated conductor, before (top) and after (bottom) they are assembled into cable.**

The assembling procedure of the Roebel cables is a rather challenging procedure, which requires a complex bending of the tapes around each other, avoiding over-bending or plastic deformation of the material. IRL solved this problem inventing innovative machineries that are able to produce such cables in long lengths [19] – see Figure 7. At KIT sample lengths up to 5 m are currently still made by hand. The advantages are the possibility to obtain a dense packing with a reduced central gap and the possibility to change cable designs, e.g. by multi-stacking of strands or cabling with additional stabilisation. In that way the potential of the cable technology can be investigated in a wider range.

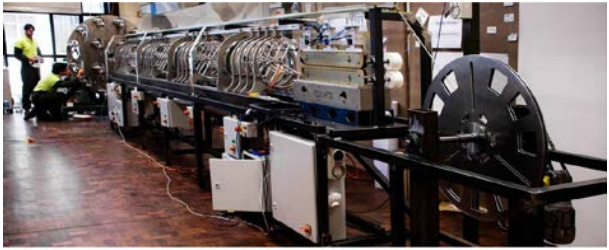

**Figure 7.** Device for assembling 15/5 (15 strands with 5 mm width) Roebel cables (courtesy IRL-General Cable).

### 2.4. Design options and modifications

A couple of cable design options exist for the different requests coming from the specific applications or operation conditions, as for example temperature and background field.

#### 2.4.1. Multi-stacking

High transport currents at 77 K can be achieved by adjusting several design parameters. Increasing the transposition length allows increasing proportionally the number of strands, resulting in enhanced critical current. The transposition length, however, needs to meet the requirements and constraints of the prospected application (coil size, loss reduction). The second way of achieving high currents is interlacing stacks of strands instead of individual strands. The potential of this method was demonstrated for both, a 12 mm-wide (3-fold stacking) [18] and a 4 mm-wide cable (3- and 5-fold stacking, see Figure 8) [22]. Stacking multiple strands is however very challenging to do automatically and so far it has been demonstrated only on cables assembled by hand. Unfortunately multi-stack cables are not very suitable for windings, because they have the general problem of stacks: the tapes in the stacks are not fully transposed and therefore the inner CCs need to be shorter to match the meander pattern of the others. For applications where limited bending occurs, for example in bus bars or straight high-current power lines, this method can be easily applied to enhance the current carrying capability of the cable. The consequences of a multi-stacking arrangement for the cable's AC losses are discussed in section 4.

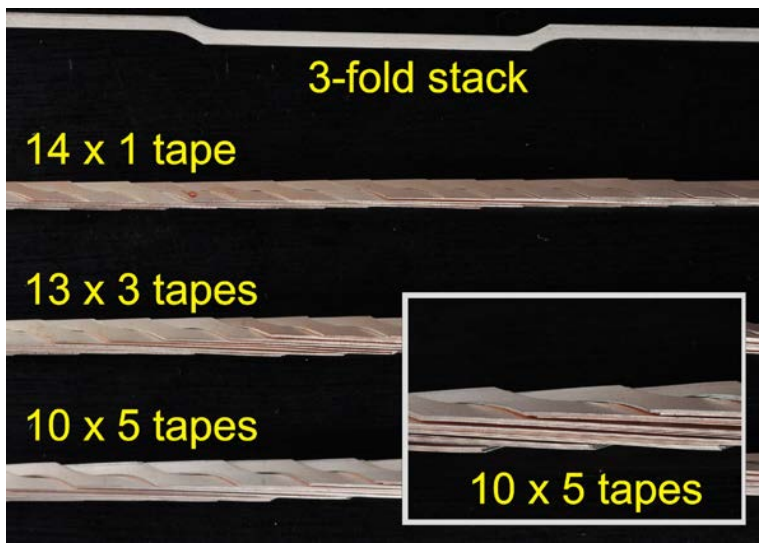

**Figure 8.** Multi-stacking of strands: a 3-fold stack of punched CC tapes and 3 different Roebel cables with 4 mm width are shown. The upper cable consists of 14 single tapes, whereas the middle and the lower cables consist of 13 3-fold stacks and 10 5-fold stacks, respectively.

### 2.4.2. Strand coupling

Usually no insulation is applied between the strands and coupling is caused by the "natural" contact between the strands. Coupling currents flow around the non-punched edges of the tapes, since the buffer layers below the superconductor are insulators. Inhomogeneous current distributions in the CC and a controlled redistribution of currents between the strands, which is desirable in low field or self-field applications, can be managed by a moderate strand coupling. One way of obtaining this coupling was demonstrated by using Ag/resin paste, which provides a moderate coupling with tolerable coupling AC losses [23]. Since the impregnated cable becomes quite stiff and is not suitable for strong bending, this option is however limited for application "in situ" where no further strong bending of the sample occurs.

Recently, Majoros *et al.* studied the effect of different kinds of inter-strand connections on inter-strand resistances, in order to increase the current sharing between the strands [24]. They found that by just applying a pressure of 8 kPa gives a maximum inter-strand resistance of 105 mΩ. Using a tin foil placed on top of the Roebel cable gives the maximum inter-strand resistance of 31.5 mΩ under a pressure of 8 kPa. The lowest maximum inter-strand resistance of 0.1 mΩ was achieved by soldering copper shunts. AC loss measurements revealed that even an inter-strand resistance as low as 0.1 mΩ does not cause any significant coupling loss increase up to 200 Hz, while allowing current sharing.

### 2.4.3. Internal and external stabilization

Industrial CCs are usually plated with stabilizing copper for thermal stabilization. The layer thickness (typically 20 μm) has to be matched with the operation current. This is particularly important when operation temperatures below 77 K are considered, where the transport currents can increase by one order of magnitude, and can lead to the necessity of using even thicker stabilization layers. Such a conductor is more difficult to be handled as Roebel strand in the cable fabrication process. In this case the solution is to apply an already meander-shaped copper tape on top of each strand, with limitations and difficulties similar to the multi-stack option. A systematic investigation of this option has not been performed so far.

Classical copper Roebel cables have completely insulated strands. The insulation technology is an important issue and the subject of patents and company's know-how. At cryogenic temperatures and in order to maintain the dense packing of the superconducting cable, impregnation techniques similar to those employed with low-temperature superconductors are the first choice. IRL realized an epoxy acrylate coating of typically 20 µm without effect on the current capacity [25]. The accuracy of the coating however is not perfect at the sharp edges of the CC. High voltage properties were not fully tested. Until now it is not clear if and where an insulation of the strands makes sense.

### 2.4.4. Current and voltage contacts

The standard technique to inject the transport current is to solder the Roebel cable into a grooved Cu-block surmounted by another copper bar. Most of the CCs require use of specific solder materials [26] to avoid thermal strains in the composite, which can lead to crack formation in the *RE*BCO layer. For shorter samples up to about 1 m length, the contact resistance dominates and determines the current distribution in the cable. A systematic experimental investigation and modelling of the current distribution in the current leads is still missing.

Applying voltage contacts for the current measurements with the standard four-point method is not trivial. Connecting all strands together leads to current redistribution at the contacts, therefore measuring a representative strand can give misleading results for the whole cable [18]. The best way is to measure all strands in parallel, which gives more information about the scattering of the properties of the strands. This can be done by placing voltage taps in different positions on the cable or behind the current leads [27]. An indirect way is to monitor the current sharing with a parallel applied copper shunt that gives an image of the *V(I)* graph of the whole cable [18].

### 2.4.5. Striations of the strands

Striations were successfully applied to Roebel strands by laser grooving (5 filaments of 1 mm width, see Figure 9) and an effect on the AC loss behaviour was observed [28–31], although further systematic investigations are needed for a deeper understanding. The applied striations however already showed the importance of sufficient conductor homogeneity and the absence of defects. By means of Hall probe scans single defects significantly limiting the local transport currents could be identified [32]. Those results confirm the considerations in reference [33] on the homogeneity criteria to select CC for the preparation of Roebel strands.

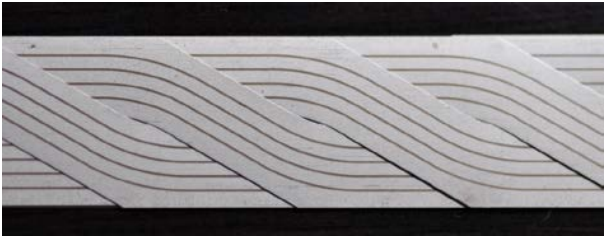

**Figure 9. Filament structure in Roebel strands made by laser grooving (KIT). Image reproduced from** [31].

### 2.5. Alternative Concepts

For HTS AC cables other alternative concepts were developed and investigated. According to the assembling geometry of HTS power transmission cables, Danko van der Laan presented the concept of the Cable on Round Core (CORC): the CCs are helically arranged around a cylindrical former in one or several layers. See Figure 10. The concept provides the twisting of the strands, it is easy in fabrication (assembling method) and is flexible in terms of current capacity (number of tapes used) [34–36]. Until now this assembly has been done by hand. A disadvantage is that the engineering current density is significantly lower than in Roebel cables. Cables with self-field critical currents of 2800 A and 7561 A at 77 K were demonstrated and first tests on smaller cables at 4.2 K and 20 T were performed [37]. Due to the special arrangement of the tapes in CORC cables self-field effects do not play a significant role and the measured critical currents of the cables are comparable to the sum of critical currents of the single tapes from which they are composed [35,36].

Another concept motivated by fusion magnet research was presented by MIT [38]: the idea is to solder 3 stacks of CCs into grooved copper rods, with twist. This cable design is known as Twisted Stacked Tape Cable (TSTC) and is shown in Figure 10. So far this concept has suffered from thermal strains in the composite. In addition, the assembling procedure of thick stacks in long lengths has not been performed yet.

In the frame of further activities to upgrade the Large Hadron Collider (LHC) at CERN, different cable concepts for MgB$_2$ and HTS CCs are currently under investigation [39]. In particular, for the CC version two concepts were proposed: a paired-CC stack approach and a layered cylindrical arrangement similar to the CORC.

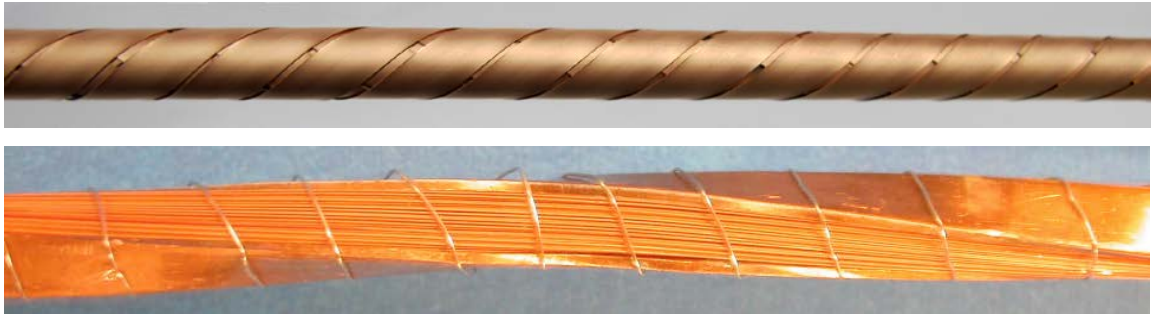

Figure 10. Pictures of two cable designs alternative to the Roebel: Top: Cable on Round Core (CORC), source: [40].; b) Twisted Stacked Tape Cable (TSTC), source: [41]. The thin wire around the CORC sample is for mechanical support during the twisting process.

## 3. Physical properties of Roebel cables

In this section the physical properties of the Roebel cable are discussed by reporting the results of DC characterization at various temperatures, of current density distributions (calculated by numerical models and obtained by scan Hall probe scanning) and of mechanical characterization.

### 3.1. DC Transport currents at 77 K

Transport currents in Roebel cables are strongly influenced by the self-field generated by the currents. A first field pattern for the cable cross section was calculated via a Biot-Savart-approach and showed a complex field distribution reaching a few hundred mT at the outer edges of the cable [18]. This quite simple approach could explain the observed significant critical current reduction caused by the self-field. Each position of the strands in the cable cross-section experiences a different field, both in terms of magnitude and direction. A finite-element method (FEM) modelling of the self-field pattern allowed a more detailed analysis, including the current anisotropy of the CC and the evaluation of the individual critical currents of the strands.

The highest transport current in a Roebel cable measured so far is 2.6 kA at 77 K in self-field, a cable with 15 x 3-fold stacked strands and a transposition length of 188 mm, made from 12 mm wide SuperPower CC. That particular cable showed some degree of current percolation between the strands, resulting in unusual current-voltage characteristics [16,18], probably to be ascribed to the fact that the copper contact was soldered on a relatively short length (about half of the transposition length). Usually, soldering the contact over one transposition length avoids the problem of those percolation currents and the current-voltage characteristics of the different strands look very similar to each other, as shown in the example of Figure 11.

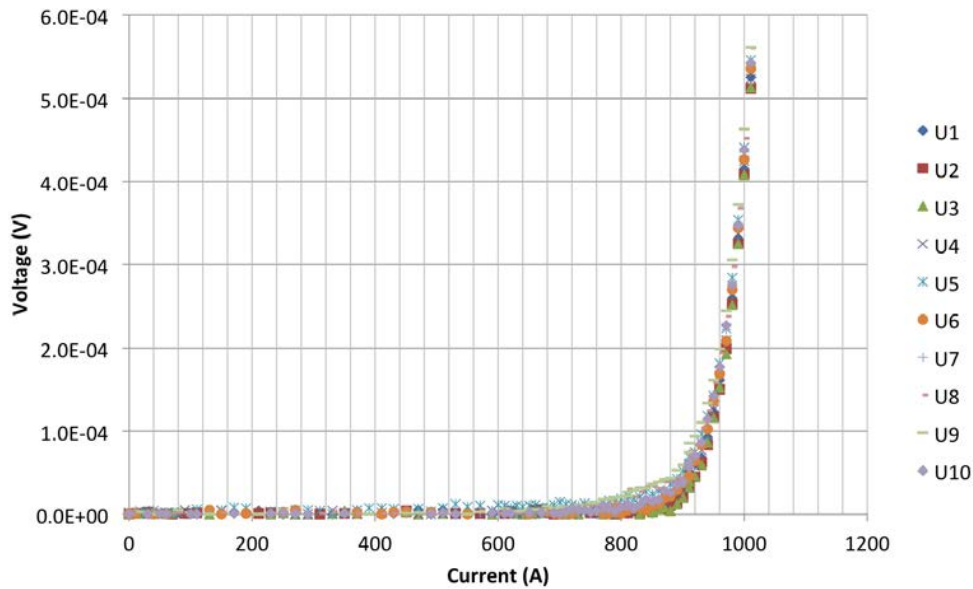

**Figure 11. Current-voltage characteristics for the Roebel cable used in** [42,43]**, measured on the 10 different composing strands. The critical current corresponding to the 1 µV/cm criterion (450 µV for a distance between the voltage taps of 4.5 m) is about 1 kA. The value of 936 A used in the references and in the table of section 3.6 refers to a lower critical voltage.**

Depending on the width of the cable (10, 12 mm or 4 mm) we can define two classes of current capacity. At 77 K the high current class can provide 1-2 kA transport current in self-field, whereas the smaller cables can provide currents in the range 0.3–0.8 kA (no strand stacking). Multi-stacking of the strands can increase the current capacity up to a factor of 2-3 with the restrictions in fabrication and application mentioned in section 2.

### 3.2. DC Transport currents below 77 K

Roebel cables are considered not only for applications at 77 K, but also for much lower temperatures. Cryocooler technology has made big progress during the last few years and offers the possibility of operating devices (motors, generators, transformers and magnets) at much lower temperatures than liquid nitrogen. Closed cooling cycles have already been tested with liquid hydrogen (20 K) and liquid neon (28 K). For bus-bars in accelerator applications helium gas cooling is considered with T=10-15 K, whereas for future fusion reactor scenarios (DEMO) HTS magnets operating at 50 K or below are being considered. CERN recently performed the first successful current capacity test in the FRESCA test facility at 4.2 K in background fields up to close to 10 T [44]. The result was a remarkable increase of the currents with lowered temperature as shown in Figure 12.

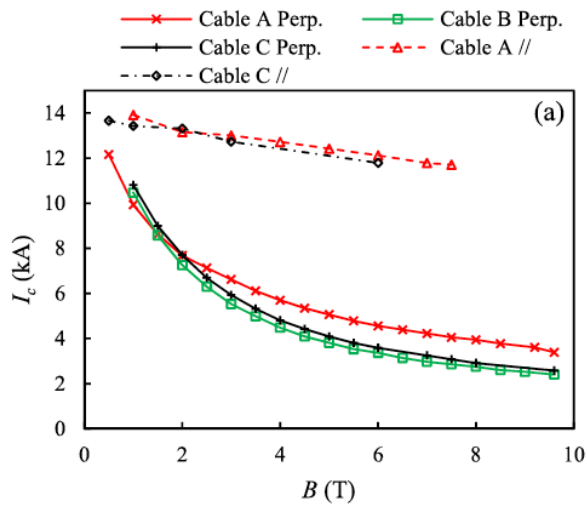

**Figure 12. Field dependence of the critical current for different cable samples at 4.2 K. Cable A is from GCS, cables B and C from KIT. Source:** [44]**.**

Conductors from KIT and General Cable Superconductors Ltd. (GCS) were successfully tested; the occurring Lorentz forces were kept under control by fixing the cables on the support by means of lateral pre-compressing. A current of about 14 kA was reached for a 12 mm-wide Roebel cable from KIT with 10 strands of 5.5 mm width at 4.2 K and a background (parallel) field of approximately 0.5 T. The field dependence was measured for both orientations of the field, perpendicular and parallel to the cable (see Figure 12). The data show a behavior very similar to the $I_c(B,T)$ dependence of a single CC tape and the increase of the transport currents from 77 K to 4.2 K is more than a factor 10. The CC material from SuperPower was significantly different for both kind of cables: BZO doped tape was used by KIT and non-doped by GCS, respectively. Although GCS used 15 strands (width 5 mm) in comparison to 10 strands (width 5.5 mm) of KIT, the achieved transport current performance was comparable, which indicates the $I_c$ of the starting materials used in the KIT cable was significantly higher on average. Due to the smaller cable cross section, the engineering current density of the KIT cable was approximately 50% higher, a big advantage for magnet applications.

In another activity pursued at KIT, the split coil magnet facility FBI was equipped with a heating jacket so that the sample's temperature can be increased beyond 4.2 K [45]. The technological problem of the heat chamber in liquid helium has not been completely solved yet, and the achieved data suffer from the fact that the sample's temperature is not precisely known [20]. The measured current values however indicate that scaling up the currents to other field and temperature conditions works quite similarly to single coated conductors. The measured results were compatible with the data obtained at CERN within the error margin.

### 3.3. Current distributions and self-field pattern

Coated conductors show strong dependence of the critical current on the magnetic field, and specifically a strongly anisotropic dependence on the orientation of the field with respect to the tape. This behaviour and also the temperature dependence of the critical current are strongly correlated with the kind and orientation of artificial pinning centres (APC) applied by doping the superconductor. The magnetic field distribution and associated local critical current density reduction inside the Roebel cable cross-section can be not trivial. A first attempt to get an insight into the behaviour at 77 K was performed with a simple model based on the Biot-Savart law [18,46]. Later, a more sophisticated analysis was carried out by means of a dedicated FEM model [47], which included the anisotropic dependence of the critical current density $J_c$ on the local magnetic flux density – see **Figure 13**: with those calculations it was possible to estimate the critical current of the strands as they move along the cable length.

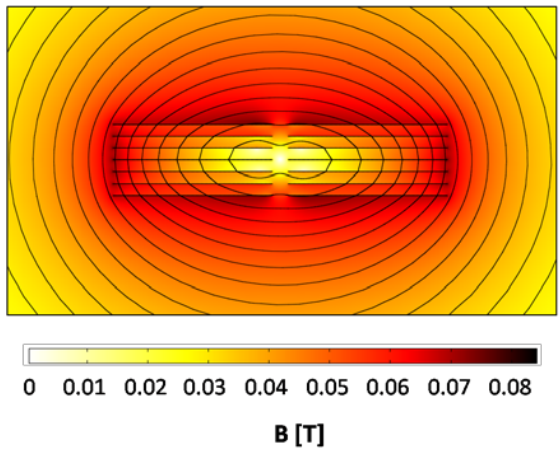

Figure 13. FEM-calculated self field pattern in the cross section of a Roebel cable. Each tape represents the different positions assumed by a tape along one transposition length. Redrawn from [47].

Scanning Hall probe investigations were performed to reveal the magnetic field pattern along the length of Roebel cable samples, which reflects the geometrical arrangement of the coated conductor strands in the cable structure [32]. The width of the space between the strands can be observed in the field map. An example is given in Figure 14. Under the action of a moving magnet, higher magnetic fields are found in the middle of the tape, although their relative size decreases as the applied field increases. Scan measurements before and after inter-strand coupling did not show significant differences in the field map.

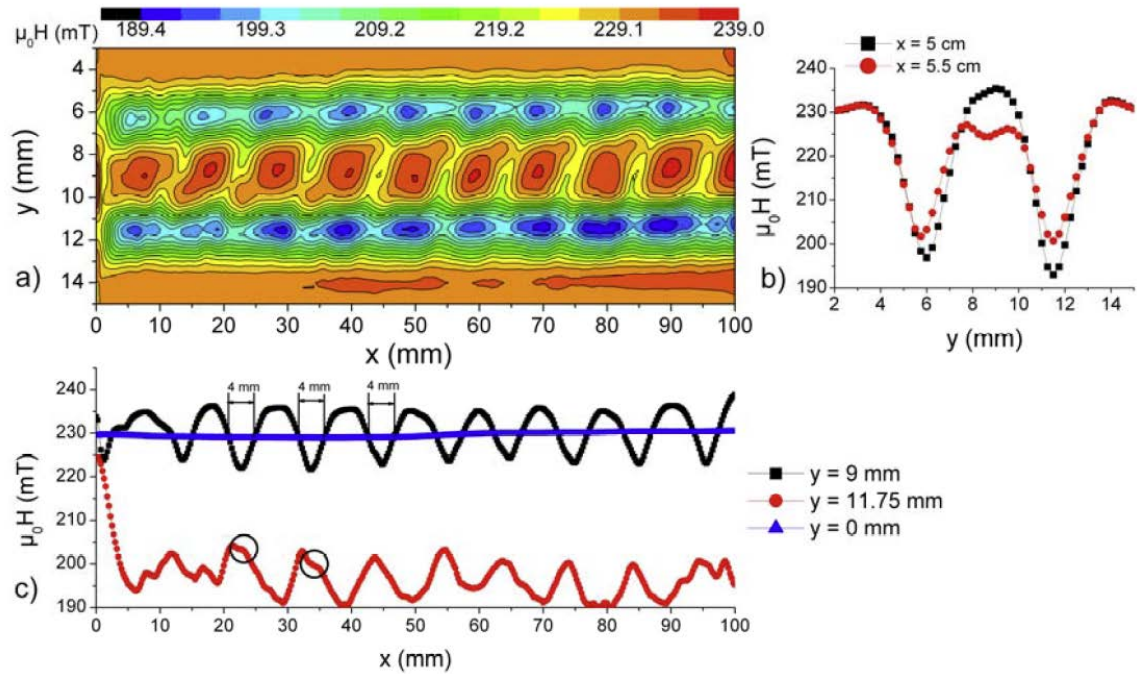

**Figure 14. Exemplary magnetic field map obtained by Hall probe scanning technique. The field profiles along different lines in the x- and y- direction are also plotted. The applied field is 230 mT. Note the different scales adopted for the width and the length of the cable in a), which make the aspect ratio of the cable different from the real one. Reprinted from** [32].

The Hall probe scanning technique was also used to investigate the magnetic field and, through the solution of an inverse problem, the current density distribution in individual strands used to assemble Roebel cables [48]. Field and current distributions in the straight part agree with the theory of thin films. A different behaviour for currents penetrating from the edges and currents penetrating from the top and bottom surfaces was found. This results in a non-parallel current flow in the crossover part of a single Roebel strand. This technique allows investigating the local properties of the conductor with high resolution and is therefore a very useful tool to find problems, especially associated with the uniformity of the conductor, in the manufacturing process of Roebel cables. Two examples of current patterns in the presence of defects are shown in Figure 15.

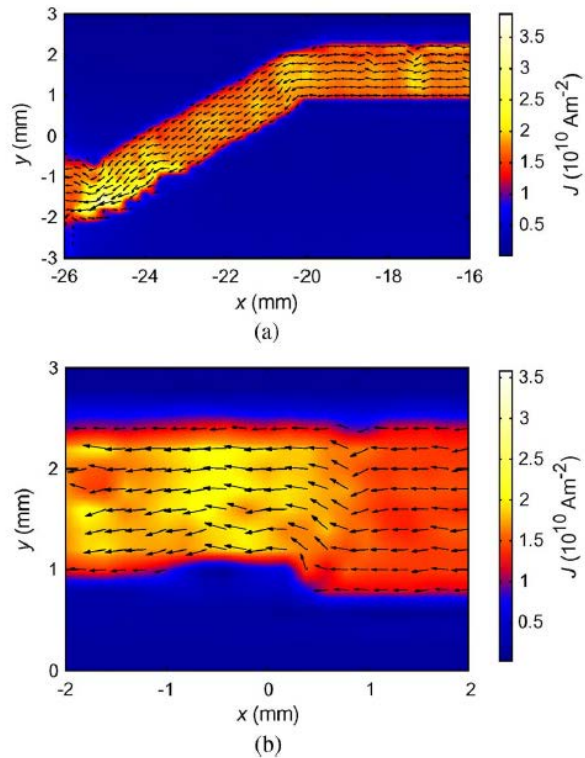

Figure 15. Two examples of current flow (reconstructed from Hall probe scan measurements) close to local defects: (a) a defect at the inner left corner leads to an enhancement of the current density close to the opposite edge (yellow – compare with the right corner, which does not present such defect); (b) due to a reduction of the cross section, the current density is increased in the left section. Reprinted from [48].

### 3.4. Mechanical properties of Roebel cables

Mechanical induced effects from applied strain have been the subject of finite-element method investigations [21,25], which identified the inner radius of the crossover section as the critical point with the peak values of mechanical stresses. It has also been found that a sharp outer edge of the crossover section, as in the KIT cables, reduces the peak strain [20].

The field dependences of critical currents of a KIT and a GCS cable were measured at 4.2 K in both magnetic field directions by Fleiter et al. [44]. In order to withstand the Lorentz forces that arouse during the measurements, the cables were fixed between two stainless steel plates. Both cables withstood transverse stresses of 45 MPa. Since the cables are not flat, peak stress areas could be identified by means of Fujifilm papers. It was found that the effective section that experiences transverse stress is quite limited and can be estimated to be only about 36% and 24% for the GCS and KIT cables, respectively. This means that a nominal loading up to an average stress of 40 MPa corresponds to a local stress of at least 111 MPa, for the GCS cable, and 167 MPa, for the KIT cables. By disassembling the cables and testing the strands, it was confirmed that no degradation of the superconductor occurred.

In another study [49], Uglietti et al. investigated the effect of transverse compressive loads on the critical current of coated conductor tape of 4 mm width, of a punched coated conductor strand 2 mm wide and of coated conductor Roebel cables (consisting of 10 punched strands). While the tapes could withstand a compressive stress larger than 100 MPa without significant degradation of the critical current (less than 2%), the degradation of the critical current of some of the individual strands in the Roebel cables exceeded 30% at a transverse compressive stress as low as 10 MPa. Visual inspection of the extracted strands revealed some local damage of the strand surface corresponding to the meander structure of the strand lying above or under the damaged strand. Subsequent measurements of the local critical current confirmed the reduction of critical current at the positions showing a damaged surface.

Bumby et al. [50] performed a similar study (supported also by numerical simulations) on tensile stresses of 12 mm coated conductor tape, 5 mm wide Roebel strands and a 15/5 HTS Roebel cable. They found that irreversible degradation of Ic occurred at 700, 146 and 113 MPa, respectively.

The apparent conflict of those results indicates that there are still many open questions on the mechanical properties of Roebel cable, which need further investigation.

Resin impregnation can help improve the resistance of Roebel cables to mechanical stresses. However, the huge difference of the thermal expansion coefficients of YBCO and resins, on top of the peculiar layered structure of coated conductors and of the meander shape of the strands in a Roebel cable, makes the impregnation of Roebel cables a quite difficult task, compared for example to the impregnation of conventional LTS cables. First attempts to impregnate a Roebel cable with Araldite and silica powder mixture are presented in [51], although the difference of the test conditions used for the impregnated and non-impregnated cables do not allow having a conclusive word on the efficacy of the impregnation technique presented there.

In view of the application of Roebel cables in high-field magnets with strong mechanical forces (hundreds of MPa), M. Bird has recently proposed the idea of a cable-in-conduit configuration, using a hastelloy jacket to provide high strength and stiffness while also providing similar thermal contraction as the YBCO tape [52].

### 3.5. Cables with very high currents

Very few possible applications of HTS cables, as magnets in fusion devices, big water power or gas turbine generators require an operation current capacity of the cable between 20 and 30 kA. In fusion magnets the projected operation temperature for HTS cables is between 4.2 K and 50 K with approximately 13 T peak field at the conductor. Within a single Roebel cable such performance is hard to achieve, but with further improved CC performance this is not impossible in the future. This performance can be reached by using multi-stacks of strands and a long transposition length, in the order of 0.5-1 m. KIT proposed a Rutherford cable design with Roebel strands as potential solution [53]. Roebel cables serve as strands wound around a central former, which can be moderately flat or round and can provide the channel for the coolant. This structure provides a transposition of the strands (Roebel cables), which are transposed as well. Efforts to realize a model cable of 0.5 m length pointed out the bottlenecks of this design [54]. The bending of the Roebel strand around the edge is a critical issue: in the experiments carried out in [54], three Roebel cables were wound onto a 10 mm thick stainless steel former with a winding angle of 20 degrees. Of the three cables, one was successfully wound with a current degradation as low as 12%; in the other two samples, the current was drastically reduced because of the solder flowing in between the strands during the soldering of the current contacts and by the subsequent bending of the stiffened area. Still open questions are the behaviour of the bent section upon cooling cycles, the need of avoiding delamination from stresses and how the cable assembling works over long lengths. An advantage is the flexibility of the design, a disadvantage the complex composite and the necessary thickness of the former (15-20 mm). In the future this concept has to compete with the potential of an upgraded single Roebel cable, already showing a self-field critical current 14 kA at 4.2 K in a standard configuration [44]. Several parameters can multiply the current capacity, for example: extended transposition, wider tapes, improved CC performance (being already shown in short samples by several producers) and finally the multi-stack strands.

### 3.6. Examples of significant cables

The following table summarizes the properties of six significant Roebel cable samples manufactured by KIT and GCS in recent years. These cables are reputed significant because of the high reached critical current (KIT-2, GCS-2) and length (KIT-3, GCS-3). Two cables manufactured with narrow 2 mm-wide strands (KIT-1, GCS-1) are also listed. In the table the design critical current refers to the value obtained by simply summing the critical currents of the individual composing strands.

|  | KIT-1 | KIT-2 | KIT-3 | GCS-1 | GCS-2 | GCS-3 |
|---|---|---|---|---|---|---|
| Number of strands | 10 | 3 x 15* | 10 | 9 | 15 | 15 |
| Original CC's $I_c$ (A) | 156 | 359 | 348 |  |  |  |
| Strand width (mm) | 1.72 - 1.82 | 5 | 5.4-5.6 | 2 | 5 | 5 |
| Strands' $I_c$ (A) | 54.3 - 71** | 149.5*** | 140*** |  | 123±1** | 105.5, 125, 180.4 |
| Transposition | 115.7 | 188 | 125.8 | 90 | 300 | 300 |

| length (mm) | | | | | | |
|---|---|---|---|---|---|---|
| Cable's length (m) | 1 | 1.1 | 5 | 0.54 | 5 | 21 |
| Cable's cross-section (mm$^2$) | 2 | 27.6 | 6 | 2.5 | 9.6 | 9.6 |
| Cable's design $I_c$ (A) | 640 | 6727 | 1400 | 427 | 1950 | 2093.5 |
| Cable's measured $I_c$ (A) | 447 | 2628 | 936 | 309 | 1100 | 1420 |
| $I_c$ reduction from self-field (%) | 30 | 61 | 33 | 28 | 44 | 32 |
| Engineering current density (A/mm$^2$) | 223.5 | 95.2 | 156 | 123.6 | 114.6 | 147.9 |
| Reference | [54] | [18] | [43,55] | [25] | [25] | [56] |

All critical currents are at 77 K in self-field.
*15 stacks of 3 strands each           ** measured           ***estimated

## 4. AC losses

Roebel cables represent an attractive solution for high-current conductors with low AC loss. Not only can they carry large currents due to the numerous coated conductor strands assembled in the cable structure, but they can assure that the current is evenly distributed between the strands thanks to their transposition, which makes each strand experience the same electromagnetic environment. This is a much more efficient solution with respect to e.g. simply stacking tapes together: in a stack of parallel conductors, the current would tend to flow in the tapes located on the top and the bottom, rapidly saturating them and causing a very high power dissipation [57]. A similar reasoning can be applied to the magnetization losses (caused by an external varying magnetic field): coated conductors suffer from very high magnetization losses in the presence of perpendicular magnetic field; stacking coated conductors may solve the problem for low applied fields (below penetration) [58], although most of the dissipation occurs on the top/bottom region of the stack [59]. Thanks to their periodically repeating and transposed structure, Roebel cables represent a possible solution to reduce the magnetization losses and equally distribute them between the strands.

In the following subsection, we review the main techniques and results concerning the measurement of transport and magnetization AC losses in Roebel cables. We also provide an overview of the main available tools to predict the loss values, including numerical models and empirical methods.

### 4.1. Experimental techniques for measuring AC losses

The techniques to measure transport and magnetization loss are very different to each other.

The methods to measure the transport AC loss in Roebel cables essentially follow the same principles used for single tapes [60]. Transport AC losses are measured by recording the voltage component in phase with the transport current, usually by means of a lock-in amplifier [61]. Due to the relatively complex geometry of Roebel cables, however, the question of how and where to fix the voltage taps arises. The most accepted solution seems to be to solder voltage taps on the top of a particular strand over an integer multiple of the transposition length [22,62] (Figure 16). This latter action is necessary in order to average the loss signal over all the positions experienced by the strand inside the cable. Jiang et al. showed in [62] that the transport loss of a Roebel cable with $N$ strands can be estimated by taking the mean value of the in-phase voltages $V_k$ measured from different voltage loops attached to the different strands as:

$$Q_t = \frac{V_{mean} I_{cable}}{d \cdot f} = \frac{\frac{1}{N}\sum_{k=1}^{N} V_k I_{cable}}{d \cdot f}$$

where $I_{cable}$ is the total current carried by the cable, $d$ is the length of the voltage loop and $f$ is the frequency.

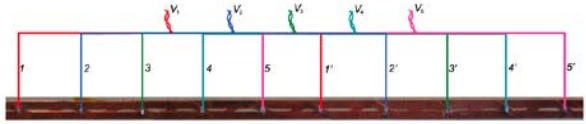

**Figure 16.** Voltage loops soldered on individual strands for measuring transport AC losses. Reprinted from [62].

Placing many voltage loops can easily become very complicated; alternatively, the voltage taps can be positioned on the current leads. The drawback is that the (usually large) resistive signal of the current leads is measured as well and needs to be subtracted from the measured voltage in order to obtain the loss in the cable. This can be quite tricky, especially because the loss signal coming from the current leads is usually larger than the signal coming from the cable. A good agreement with the "standard" approach with the taps soldered on a strand has been reported in [22], although the authors acknowledge that it could be coincidental and deserves further investigation. Another situation where the placement of voltage taps on a strand could be difficult is in the case of coils, especially if they are tightly wound [42,63].

Measurement of magnetization AC losses is typically performed by placing a short piece of Roebel cable (of length at least equal to the transposition length) in a uniform magnetic field usually generated by coils. Losses are then determined either by measuring the magnetization of the sample by means of pick-up coils [64,65] or by means of the calibration-free method [66]. Measuring a short sample of cable, however, presents the problem that the magnetic flux can enter from the ends of the cable; for this reason, experimental results evaluating the uncoupling between strands (or filaments) should be carefully evaluated in this perspective.

### 4.2. Modelling of AC loss

Estimating the AC losses in a Roebel cable is not a very easy task, mainly because of the meander-like shape of the strands and their braid arrangement in the cable structure.

The simplest approach to model a Roebel cable (both in the case of using analytical expressions or building a numerical model) is to consider only the transversal (2-D) cross-section of it. This simplified approach is based on the fact that the transposition length is much longer than the strands' width, and the currents crossing the cable's whole width are expected to have negligible effects – see schematic representation in **Figure 17**.

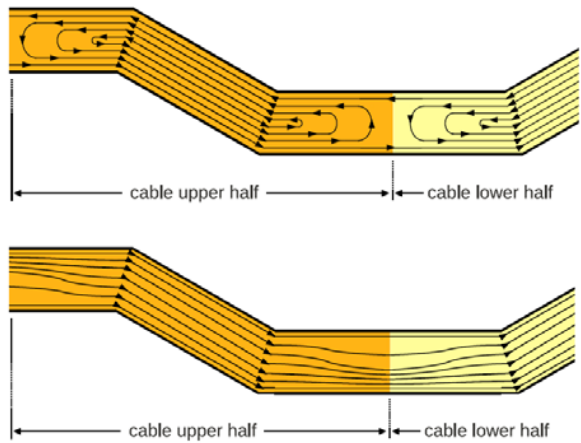

Figure 17. Qualitative representation of the current paths in a Roebel strand in a multiple-strand cable submitted to an applied magnetic field (top) or a transport current (bottom). The longitudinal current density changes along the strand length by means of a current density component in the width direction. In general, the component of the current density in the tape width is much smaller than the longitudinal one because the transposition length is much larger than the strand width. Reprinted from [67].

With this geometrical simplification, analytical expressions can be used for having an approximate estimate of the loss. In particular, for applied fields much larger than the cable self-field, one can use Bean's slab model [68] paying attention to use the proper value for the slab width $w$ used in the model, i.e. the strand or cable width for the uncoupled and coupled cases, respectively. This model also assumes constant critical current density $J_c$ and the sharp $E(J)$ relation of the critical state model [69]. The AC loss per cycle and cable length in this case is $Q = \mu_0 d w_c^2 J_{c,coupled} H_m$ and $Q = 2\mu_0 d w_s^2 J_{c,uncoupled} H_m$ for the coupled and uncoupled cases, respectively, where $w_c$ and $w_s$ are the cable and strand width, respectively, $d$ is the total cable thickness, $J_{c,coupled}$ is the engineering $J_c$ of the whole cable and $J_{c,uncoupled}$ is the engineering $J_c$ of one column of strands in the cross-section (see **Figure 13**). This equation suggests that the Roebel cable transposition reduces the AC loss by a factor around 2, since the strand width is around half of that of the whole cable and the horizontal gap between strands in **Figure 13** is small. However, the equation above is not applicable for magnetic-field dependent critical current densities, unless the cable is submitted to a DC magnetic field much larger than the amplitude of the AC magnetic field.

The Norris strip and ellipse formulas [70] provide a rough estimation for the transport loss, assuming a constant $J_c$.

$$Q_E = \frac{\mu_0 I_c^2}{\pi}\left[(1-i)\ln(1-i)+(2-i)\frac{i}{2}\right] \quad \text{(ellipse)} \quad (1)$$

$$Q_S = \frac{\mu_0 I_c^2}{\pi}\left[(1-i)\ln(1-i)+(1+i)\ln(1+i)-i^2\right] \quad \text{(strip)} \quad (2)$$

where $i=I_a/I_c$ is the normalized amplitude of the applied current $I_a$.

The analytical models described above provide however a very rough description of the physics and geometry of the cable. More realistic descriptions of Roebel cables can be obtained by means of numerical models. Most models still describe the cable as two stacks of rectangular tapes, but they have the possibility of including the various parts composing a coated conductor (e.g. substrate, stabilizer) and – more importantly – of controlling the current distribution between the strands. Typically, two extreme situations can be considered: one can leave the current free to distribute between the strands (as if they were electrically in parallel) or force an even current distribution between them. This second condition corresponds to what is expected to happen in reality thanks to the transposition of the strands and results in lower losses [22]. When only an external AC magnetic field is applied, these two situations correspond to complete coupling or uncoupling of the strands. Which of these two situations is most beneficial from the point of view of losses depends on the magnitude of the applied field [22], the geometry of the cable (e.g. cable thickness, gap), and the direction of the applied field. For all cases, at high applied fields the uncoupled case presents the lowest loss (see **Figure 18**). The behaviour at low applied fields depends on its orientation. For parallel applied fields, the uncoupled case is still favourable [71], while for perpendicular ones the coupled case presents lower losses [59]. For the latter situation, the geometry plays an important role. The narrower the horizontal gap, the wider the peak in the loss factor for the uncoupled case (and the higher the loss at low applied fields) [72]. With decreasing the cable thickness, this effect becomes more pronounced.

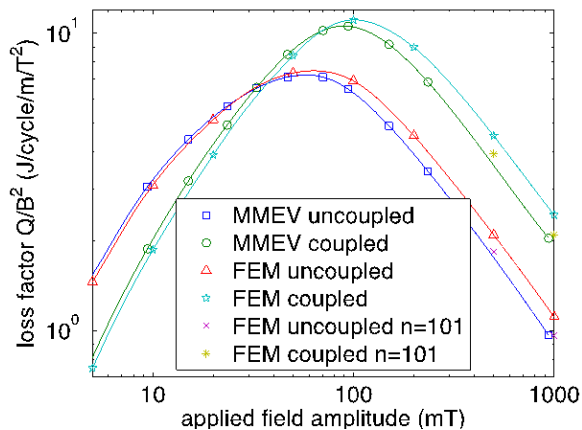

Figure 18. Comparison of the magnetization loss in a 14-strand Roebel cable for the cases of full coupling and full uncoupling between the strands, calculated with the power-law and critical state models, solved with finite element (FEM) and Minimum Magnetic Energy Variation methods (MMEV). Reprinted from [67].

The simulation of Roebel cables as two stacks of coated conductor was first proposed in [22], and further refined by Pardo and Grilli in [67] and [73]. Those publications confirmed the results discussed above and in particular [73] showed the importance of the parallel component of the magnetic field. More specifically, the AC loss depends on the parallel component of the applied field for two reasons. First, at low applied fields and low angles, the AC loss due to the penetration across the thickness is important; second, because for YBCO the field anisotropy in $J_c$ is relatively weak. Then, $J_c$ reduces significantly with increasing parallel component of the applied field for the same perpendicular component of the applied field. This is very promising for solenoids, where the AC loss due to the parallel component of the applied field is important. The transposition of the strands has only a moderate effects in reducing the transport losses, as confirmed by simulations and experiments – see Figure 20 and section 4.3.1, respectively.

Numerical models have been extended to calculate the losses in coil geometries, by means of axis-symmetric models [42,71]. In [43] the authors added a dedicated 3-D model for estimating the losses in the copper contacts as well. 2-D simulations have also been used to calculate the field profiles in cable made from tapes with magnetic substrate [74].

The first 3-D model for a Roebel cable was proposed by Nii et al. [75]. The model is for example able to calculate the current density distribution and the local dissipation in the crossing points (see **Figure 19**). That work has been recently extended in [76], where the authors come to the conclusion that while the current lines on the strand of the Roebel cable and the mode of magnetic flux penetration are influenced by the three-dimensional transposed structure of the Roebel cable, giving rise to regions of high loss concentration, the AC loss of the Roebel cable averaged along its transposition length is almost the same as those of the stacks of coated conductors, which simulate bundled conductors with uniform current distribution. For sake of precision, it has to be mentioned that the model presented in [75,76] considers the superconductors to be infinitely thin, and as such it cannot take into account the magnetic flux penetration from the wide faces of the superconductor. This limitation has been recently overcome by Zermeno et al. [77], who have proposed a fully 3-D model for a Roebel cable: while the model is able to solve a fully 3-D problem, its routine use may suffer from the intense computational effort and issues related to mesh generation and optimization.

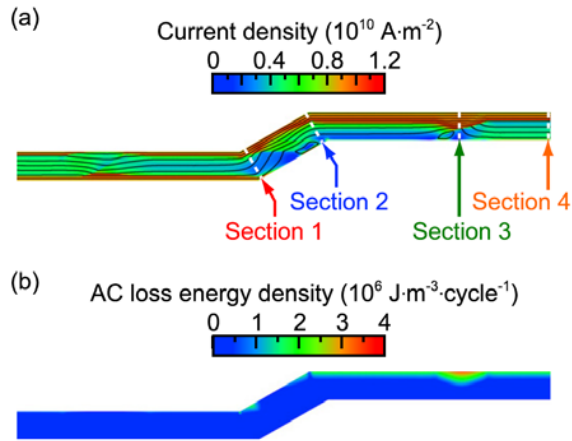

Figure 19. Current density (a) and AC loss density (b) distributions in a six-strand Roebel cable carrying AC current and subjected to an AC perpendicular magnetic field. Reprinted from [75].

### 4.3. Observed AC loss behaviour

In this section we report the main experimental results on transport and magnetization loss measurements on Roebel cables. In the following, we focus on cables made from coated conductors with non-magnetic substrate because of the dwindling use of magnetic substrates in applications, especially in view of recent findings which indicate the possibility of utilizing substrates with virtually non-magnetic behaviour for manufacturing coated conductors with RABiTS technique [78]. At the end of this section, we present a brief summary of the AC loss behaviour of Roebel cables made of tapes with magnetic substrate and we list the relevant references.

#### *4.3.1. Transport loss*

Jiang et al. measured the transport loss of a Roebel cable composed of five strands with non-uniform current distribution in a range of frequency from 59 to 354 Hz [62]. It was found that the losses are mostly hysteretic and that the average voltage method described above is a valid estimation of the losses of the cable. A mostly hysteretic loss contribution was also reported in [22], where the transport loss of cables composed of individual strands and stacks of strands was measured.

Numerical simulations based on an equally distributed current between the strands agree well with experimental values.

The spacing between the strands ($W_c$ parameter in Figure 4) influences the magnetic field experienced by the superconductor, and therefore the AC loss of the cable. In particular, a larger spacing increases the critical current and decreases the transport losses, so spacing the strands may be useful for reducing the cable loss in applications where transport loss is dominant, although this reduces the cable's engineering current density [79].

Evaluating the impact of the Roebel geometry on transport loss characteristics is not straightforward, because different objects can be used for comparison. For this purpose, Norris's formulas can be of help.

One can compare the loss of a Roebel cable composed of *N* strands with the loss of *N* conductors bundled together, as done in [62]. In that case, the loss of a conductor in a bundle should be *N* times higher than the loss of an isolated conductor. However, experimental findings in [62] indicate a smaller loss increase. In a later work [80] the authors compare the transport loss of a Roebel cable composed of eight 2 mm wide strands with a stack composed of four 4 mm wide tapes connected in series to impose the same current in each tape: at low currents the loss is similar, but at medium-high currents the Roebel cable has lower losses; in particular, at $I_t$=0.99$I_c$ the loss of the Roebel cable is about 30% lower than that of the stack. However, more than as a positive effect of the transposition, this can be seen as a simple effect of the central gap inside Roebel cable: the strands in the Roebel cable are more loosely arranged than in the stack (where there is no gap), the self-field is lower and the losses are consequently lower as well. This kind of AC loss reduction has been discussed for two tapes in [81] (see figures 4 and 5 of that article).

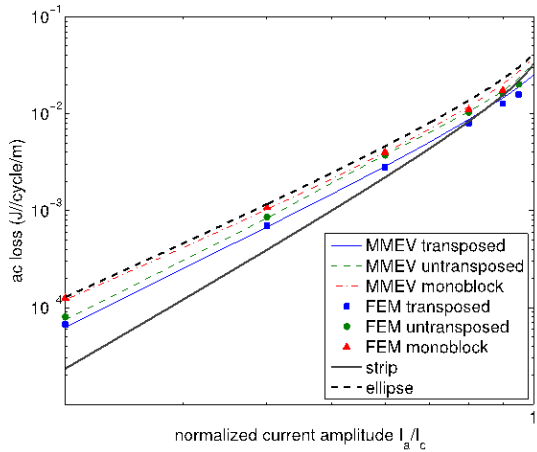

**Figure 20. Transport AC loss for transposed and untransposed strands with a cross-section representative of that of a Roebel cable (two stacks of tapes), calculated using two different models (FEM and MMEV, refer also to** Figure 18**) and compared to the loss of a rectangular monoblock. Figure reprinted from** [67]**.**

Alternatively, one can compare the loss of a Roebel cable to those of conductors (with elliptical or rectangular cross-section) with the same critical current of the Roebel cable. In [22] it was found that the measured loss falls between that of an elliptical and thin rectangular conductor. Numerical calculations for two stacks of tapes with uniform current distribution among them agree well with experiments. For the same Roebel cable geometry, in [67] the authors compared the calculated transport loss for the transposed and untransposed case, finding that the transposition reduces the loss by about 20% (Figure 20). This is due to the fact that the Roebel cables under study are thin objects, and for such geometry the current is already approximately balanced between the strands even without transposition; therefore, the transposition can reduce the loss only moderately.

### *4.3.2. Magnetization loss*

The first measurements of magnetization loss were reported in [82,83]: a weak frequency dependence was observed, which indicates that the most important loss contribution comes from the superconductor losses. An even weaker frequency dependence was observed on the samples with different architecture presented in [22]. In that paper no special resistive material was used for reducing the coupling loss; this means that the contact resistance between strands is sufficiently high for suppressing the coupling loss. In [84] Lakshmi et al. ascribe the observed slight frequency dependence of the loss to the intrinsic frequency dependence of the loss of the superconductor, and they use finite-element method (FEM) calculations to support this conclusion – see also [85], [86].

In order to increase the current carrying capability, Terzieva et al. [22] prepared cables with stacks of 3 and 5 strands. This leads to a better (partial) shielding of the applied magnetic field, to the shift of the field of full penetration to higher amplitudes, and to the reduction of the loss at low applied fields.

In [87] the authors compared the magnetization loss of a 5-strand Roebel cable with insulated strands to that of an individual strand, showing a reduction of the loss at low and intermediate fields and a convergence of the two loss curves at higher fields, when the penetration of the cable is complete. They also show that the angular dependence of loss obeys a simple scaling law, according to which only the perpendicular component of the magnetic field matters [87]. It has to be remarked, however, that the loss results presented there span over six orders of magnitude, so variations in the order of 20-30% are scarcely visible. In fact, calculations carried out by Pardo and Grilli [73] showed a more complex behaviour caused by the complex angular dependence of $J_c$ on the magnetic field. They found out that, while the dependence of loss only on the perpendicular component of the magnetic field is a reasonable approximation for angles higher than 15-30 degrees (perpendicular field corresponding to 90 degrees), this is no longer the case in general, as can be seen from the plot of the loss function $Q/H_m^2$.

In [87] the authors manufactured a coupled cable (which is more stable against defects) by coupling the strands by means of Cu-bridges: the coupled cable showed the dominance of coupling currents at low frequencies and a saturation of loss at high field, a situation where the cable behaves as a monolithic conductor with higher loss than that of an uncoupled cable.

In [87] Lakshmi et al. analyse the frequency dependence of the magnetization loss for two types of Roebel cables: an "uncoupled cable" (where the strands are electrically insulated from each other) and a "coupled cable" (where the strands are not individually insulated and some degree of coupling is present). A very weak frequency dependence of the magnetization loss is reported for the uncoupled cable: losses decrease with frequency at low fields and increase with frequency at high fields, the cross-over between the two behaviours occurring approximately at the peak of the loss function. This is due to the intrinsic frequency dependence of the superconductor material due to the flux creep, as confirmed by analytical [88] and numerical [86,89] calculations. The same dependence is also described in [90], where the two frequency regimes and their cross-over is more visible, and where an additional loss contribution is reported. Overall, for the uncoupled cable, the frequency dependence is in the range of 10% in one decade of frequency, and hence negligible for practical applications. A much stronger dependence is observed for the coupled cable, especially at low fields, and ascribed to the coupling between the strands; at higher fields the coupling contribution is less important and the loss curves for different frequencies converge

Roebel cables with striated strands have been assembled. The strand striation further reduces the loss [28,30,31,91] see **Figure 21**; however, the effectiveness of this method for reducing the loss in practical application has still to be proven, because in the experimental set-up: the filaments are electrically isolated from each other at the ends, and hence there are no coupling currents, whereas in real situations the filaments will be coupled at the ends. The concept of additional transposition by means of the more complex structure of a Rutherford cable with Roebel cables used as strands was proposed in [53] and tested with applied Roebel strands recently [54]. However, striated strands were not investigated in that work.

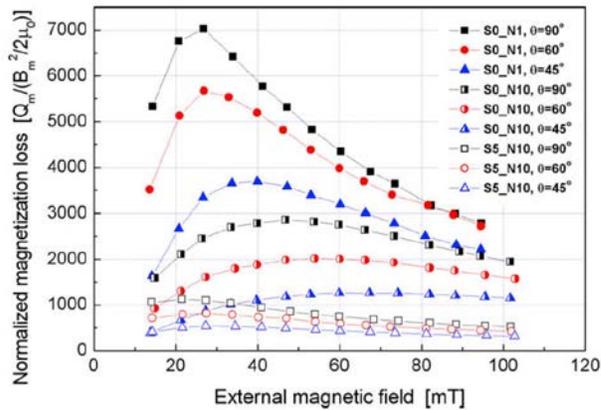

Figure 21. Reduction of magnetization loss due to Roebel cabling (S0_N10) and successive filamentarization (S5_N10) with respect to the loss in the original coated conductor (S0_N1). The angles represent the orientation of the magnetic field with respect to the flat face of the cable. Reprinted from [30].

### 4.3.3. Combination of transport and magnetization loss

The only report on AC losses for the case of combined transport and magnetization is a recent paper from Jiang et al. [92]: the author measured the AC loss resulting from the combination of transport current and magnetic field with varying orientation with respect to a tape made of six 2-mm wide strands. The AC loss characteristics are quite complicated, due to the interaction of transport and magnetization losses, and the different possible orientation of the external fields. In brief, the results can be summarized as follows. When the field is almost perpendicular to the flat face of the cable, the losses scale with the perpendicular component of the field and increase with the transport current. As the field becomes more parallel, the relative weight of the magnetization losses decreases, and the losses are mostly due to the transport current. In that paper, the authors provide a useful maximum entropy model that fits the losses in the presence of transport current and perpendicular magnetic field, and later quickly evaluate them.

### 4.3.4. Cables with magnetic substrate

For the convenience of the reader, we list here the main results of AC loss measurements on Roebel cables assembled from tapes with magnetic substrates: The observed large transport loss is due to the additional contribution of the substrate; in addition, electromagnetic calculations indicate that most of the substrate loss occurs in the strands situated at the bottom of the cable's cross-section [74]. In perpendicular applied field, the magnetization loss is dominated by the substrate loss at low field amplitude [93]. In parallel field, the substrate loss is the dominating component; the unexpectedly large observed frequency dependence is not intrinsic to the substrate, and has been ascribed to current loops inside the strands (coupling the superconductor layer and the copper stabilizer) enhanced by the magnetization of the substrate [90].

## 5. Summary and outlook

In the past few years, the HTS Roebel cable has been investigated by different groups around the world, particularly by KIT in Germany and IRL and GCS in New Zealand. This cable concept matched the expectations in terms of high current-carrying capability and reduced AC losses. In view of AC applications, it is the only HTS cable concept that provides full transposition of the strands with a compact cable design that leads to high engineering current densities. The cable design offers different options to increase the transport current, the stabilization and the geometry. Roebel cables are currently commercially available, which is leading to an increased activity on possible applications. A unique feature of this cable is its good bending capability, which favours coil applications as stator and rotor windings for rotating machinery but also transformers and magnets. In the latter case, low temperature operation at 4.2 K or in the range above is of extraordinary interest in virtue of the impressive increase of the current-carrying capability of *RE*BCO coated conductors (exceeding one order of magnitude with respect to 77 K). HTS Roebel are currently being considered as insert coils in dipole magnets of accelerator facilities and in high-field magnets exceeding fields of 40 T: for this kind of applications and magnetic fields, the tolerance to the mechanical stresses has still to be conclusively proved. Roebel cables from the mostly advanced coated conductors have the potential to be scaled up to a current capacity of more than 20 kA at fields around 13 T and temperatures between 4.2 K and 50 K, which is the requirement for fusion magnets of the next generation (DEMO). A substantial penetration of Roebel cable in the market will be possible thanks to the increase of the production speed and to the decrease of the coated conductor material, and to technological breakthroughs such as increased design flexibility. A further possible step to reduce the production costs is an advanced approach that starts the *RE*BCO coating on an already meander-shaped Roebel substrate. The expected increase in demand and further development of Roebel cables may drive the manufacturers to follow this route.

## Acknowledgements

The authors would like to thank Nick Long from IRL for providing unpublished material. F. Grilli and M. Vojenciak acknowledge financial support from the Helmholtz Association (Helmholtz-University Young Investigator Group Grant VH-NG-617). E. Pardo acknowledges financial support of Euratom FU-CT-2007-00051 and the SF of EU from the Ministry of Education, Science, Research and Sport of Slovakia (ITMS 26240220028).

This is an author-created, un-copyedited version of an article accepted for publication in *Superconductor Science and Technology.* The publisher is not responsible for any errors or omissions in this version of the manuscript or any version derived from it. The Version of Record is available online at http://dx.doi.org/10.1088/0953-2048/27/9/093001